# Wireless Home Automation Using Social Networking Websites

Divya Alok Gupta
Texas Instruments Pvt. Ltd.
Bengaluru, India
divyaalok@ti.com

C.Y.N. Dwith
Texas Instruments Pvt. Ltd.
Bengaluru, India
cyndwith@ti.com

B. Aditya Vighnesh Ramakanth
Texas Instruments Pvt. Ltd.
Bengaluru, India
aditya.vighnesh@ti.com

*Abstract*— With the advent of Internet of Things, Wireless Home Automation Systems (WHAS) are gradually gaining popularity. These systems are faced with multiple challenges such as security; controlling a variety of home appliances with a single interface and user-friendliness. In this paper we propose a system that uses secure authentication systems of social networking websites such as Twitter, tracks the end-user's activities on the social network and then control his/her domestic appliances. At the end, we highlight the applications of the proposed WHAS and compare the advantages of our proposed system over traditional home automation systems.

## I. Introduction

WHAS systems described by Alkar et al.[1] and Nunes et al.[2] transfer control from wall mounted power switches to an on-screen graphical user interface (GUI). The role of the end-user is limited to selecting appropriate switches to drive his/her household appliances. Such systems allow the user to control the on/off capability and select particular modes of operation for various domestic appliances. Moreover, these systems also provide the flexibility to create custom interface buttons/switches that configure certain household appliances. For instance, if the user wants to switch on a light and set its intensity, he/she must select the respective button in the WHAS's GUI, switch it on, and then use the appropriate knob to adjust the intensity. Alternatively, a custom button can be created on the GUI that switches on the light and configures its luminescence to an arbitrary pre-programmed value. While using home automation systems governed by this concept, the user must remember pre-configured combinations of every custom or generic switch on the interface. The effort involved in memorizing this data outweighs the benefits. In addition, WHAS have always been plagued with issues related to security. The idea of remotely controlling electronic appliances at home seems very attractive, but if this remote access control falls in the wrong hands, it can spell disaster[3].

To tackle the above described challenges and many more[4], an inordinate amount of research is being conducted in the fields such as gesture recognition[5,6] and speech/speaker recognition[7]. These technologies can potentially add a layer of security and simultaneously provide an intuitive method of conveying the user's intent to the home automation system. In this paper, we present an approach to control multiple domestic appliances through social networking websites. Social networking on the web has proliferated into everyone's life today. Tweeting and updating status online have become spontaneous activities for the modern and tech-savvy human being. In addition, social networking websites are trying their best to maintain privacy of the uploaded content and secure personal information of all users. Our method therefore, leverages both the intuitive interface and the inbuilt security features of these websites to impart a secure and user-friendly interface for home automation.

The paper is organized as follows: Section II provides a bird's eye view of the various sub-components in the system. Details on hardware design, implementation and software are provided in Section III. The last section describes the scope of research and development work required to enhance the practical feasibility of the proposed system.

## II. System Overview

The proposed system comprises three of sub-systems,

1. A sub-system to observe and gather data about the user on social networking website.
2. Multiple wireless nodes that control all the home appliances.
3. A central hub that acts as an interface between the internet and the local network of wireless nodes.

To demonstrate a use-case, we present a low cost WHAS which accesses the user's tweets and then controls his/her household appliances.

## III. Implementation

As mentioned in the previous section, the three sub-systems in our use-case are

1. A web-server to get data from Twitter.
2. Sub-GHz radio frequency based wireless nodes
3. A central hub containing both Wi-Fi and RF radios.

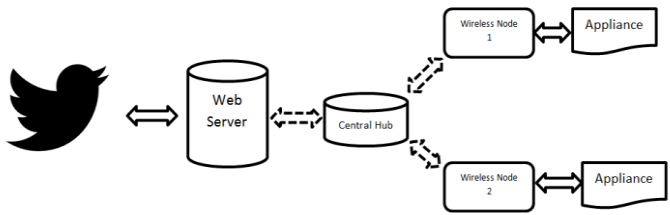

**Figure-1 Implementation Overview**

### A. Web Server

The system requires communication with Twitter in a highly secured manner. Only high-end microcontrollers, like Arduino Yun [8], are capable of handling the memory and processing requirements of such encryption/security. In order to reduce BoM for the system, an ultra-low-power micro-controller was used in the central hub while the overhead of the encryption processes was transferred to a separate web server.

The web server performs the following tasks:

- Runs a Twitter application that gets the latest tweets issued by the user.
- Runs the tweets through the Natural Language Processing (NLP) algorithm.
- Matches the results of the NLP algorithm to a look up-table and conveys appropriate control signals to the central hub.
- Renders a web application for manual control.

The Twitter application communicates with Twitter by first authenticating itself using the secure OAuth 2.0[9]. OAuth is an open standard for authorization, used to authorize third party access to Google, Facebook or Twitter accounts, without disclosing the access credentials. Once the authentication is done, the Twitter application uses REST API v1.1 [10] to get the latest tweets by the user. The obtained tweet is then processed by a decision tree based natural language processing (NLP) algorithm to obtain the user's intent. In addition to the intent, the tweet is also analyzed to get the sentiment/mood of its author. The sentiment analysis carried out by using online Twitter Sentiment Analysis API[11], is basically a fail-safe option, which can be used if and when the NLP algorithm is unable to generate usable output from the user's tweet. Based on language analyses results certain control words are generated which are sent via TCP/IP protocol to the central hub.

### B. Wireless Nodes

The purpose of a wireless node is to interface with an individual home appliance and control it. The control logic embedded in such nodes is specific to the appliance it is connected to. For instance, if a node is connected to a light system (LED/tube/fluorescent) it should be capable of handling the intensity of the light; however the same node cannot be used to control the window blinds since that would require it to control motors that open/close the blinds. Besides control, the wireless node should be able to communicate the status of the appliance such as intensity, on/off, etc. to a central hub through a wireless channel. For this use-case demonstration we used 2 wireless nodes, each controlling a set of lights and fans, as shown in Figure (1).

The wireless nodes comprise of the following hardware components:

- Texas Instruments MSP430G2553 Launchpad [12]
- Anaren CC1101 Air Module booster pack [13]

The wireless nodes communicate using the AIR RF protocol developed by Anaren[14]. These RF radios operate in the sub-GHz band. In this use-case, sub-GHz RF communication was employed over other popular RF protocols[15] such as Zigbee, because radios operating in sub-GHz band possess higher wireless range and a simpler software stack which can be used with the MSP430 family of micro-controllers. The details of operation for the wireless nodes are explained in Figure (2). In the beginning, the Master/Hub Node broadcasts an initialization command to all slave nodes. Each slave node responds by initializing the appliances (it controls) and sending its node address back to the hub. It also returns the status of control signals used to drive the household appliances. This initialization helps in identification of all slave nodes within the communication range of master node. Additional nodes can be added to the network by simply re-initializing the master node. Following the initialization phase, the master hub sends appropriate control signals to the slave nodes and receives acknowledgement from each of them.

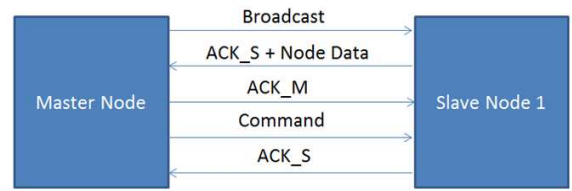

**Figure-2 Wireless Master-Slave Node Communication**

### C. Central Wireless Hub

The central hub acts an interface between the wireless nodes and the internet, which in this case are the RF nodes and web server. The purpose of the hub is to retrieve control information from the web server and then relay the same to the appropriate slave node.

The centralized Hub contains

- CC3000 Wi-Fi module[16]
- Anaren CC1101 Air Module booster pack
- MSP430F5529 Launchpad[17]

The CC3000 Wi-Fi module is used to connect to the web via the TCP/IP protocol. The ability of the CC3000 to connect to secure Wi-Fi networks via Texas Instruments' SmartConfig

[18] App makes it an ideal choice for this demonstration. The microcontroller and Wi-Fi module together act as an HTTP client which can retrieve data from the web server. Figure (3) illustrates the communication process with web server.

Along with the Wi-Fi module, the microcontroller is also connected to an RF module that acts the master node. It is connected to multiple slave nodes in a star-topology. Hence, only the master node is capable of sending or receiving data from any node.

The microcontroller and RF network software flow is described in Figure (4). In the beginning, all slave nodes are initialized using a broadcast message. The master then sequentially receives slave node addresses as part of acknowledgement message transferred by each slave. After successful acknowledgement of initialization, the master relays the control signals obtained from the web server to the slave nodes one at a time. In case of 2 slave nodes, the master sends control signals to the first slave node and waits for acknowledgement. Once it has received acknowledgement from the first node, it repeats the process for the second node.

The microcontroller uses Serial Peripheral Interface (SPI) to communicate with Wi-Fi and RF modules. It first enables the Wi-Fi Radio, gets data via the TCP/IP stack and disables it. Then, it enables the RF radio and sends out control signals for the slave nodes. Once the desired signal transmission is over and handshaking signals have been exchanged, the RF radio is also disabled.

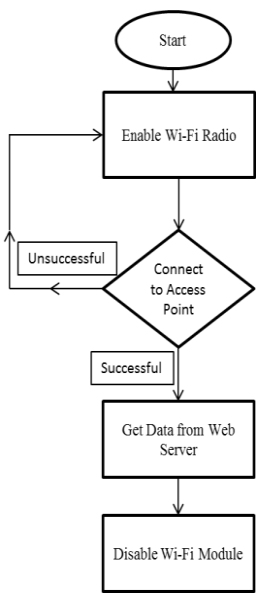
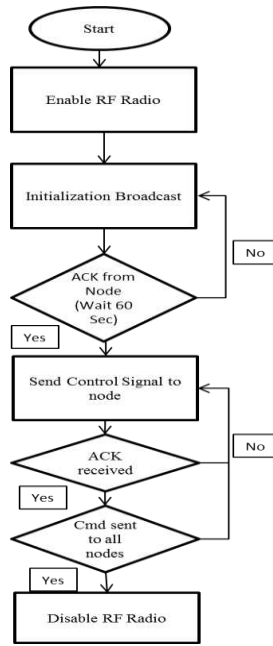

**Figure-3 Wi-Fi Flow    Figure-4 Sub-GHz RF Flow**

## IV. CONCLUSION AND FUTURE WORK

The paper outlines the development of a wireless home automation system, in which domestic appliances can be controlled by a social networking platform such as Twitter. This model of home automation liberates the end-user from the mundane task of pressing wall mounted switches or selecting them on a graphical user interface, as is the case with traditional home automation systems. The system relies on the internal security of the social networking platforms, and hence is as secure as the social networking system. Apart from this, the system can be enhanced to include speech to text or handwriting recognition engines in form of Android/Windows/iOS apps which offer another layer of security and can make home automation systems even more intuitive and convenient to use.

## *Acknowledgement*

The system described in this paper was demonstrated at Texas Instruments India Educator's Conference 2014 organized by TI India's Pragati team. We express our gratitude to Dr. C P Ravikumar and the entire Pragati team for providing us with all the support to make this project a success.